\documentclass[aip,apl,twocolumn,superscriptaddress,reprint]{revtex4-1}
\usepackage{graphicx}
\usepackage{epsfig}
\usepackage{amsmath}
\usepackage{amsfonts}
\usepackage{amssymb}
\usepackage{fouriernc}
\newcommand{\NTT}{NTT Basic Research Laboratories, NTT Corporation, 3-1 Morinosato-Wakamiya, Atsugi, Kanagawa, 243-0198, Japan.}

\newcommand{\KYOTO}{Institute for Chemical Research, Kyoto University,
Gokasho, Uji-city, Kyoto 611-0011, Japan}

\newcommand{\keio}{Faculty of Science and Technology,
Keio University, Hiyoshi, Kohoku-ku, Yokohama 223-8522, Japan}
\newcommand{\AIST}{Correlated Electronics Group, Electronics and Photonics Resarech Insitute, National Institute of Advanced Industrial Science and Technology, Higashi, Tsukuba, Ibaraki 305-8565, Japan}

\newcommand{\ket}[1]{\ensuremath{\vert#1\rangle}}
\newcommand{\bra}[1]{\ensuremath{\langle #1\vert}}

\newcommand{\beq}{\begin{equation}}
\newcommand{\eeq}{\end{equation}}
\newcommand{\beqa}{\begin{eqnarray}}
\newcommand{\eeqa}{\end{eqnarray}}

\newcommand{\cre}[1]{\ensuremath{#1^\dagger}}

\newcommand{\ddt}[0]{\frac{\mathrm{d}}{\mathrm{d}t}}

\begin{document}
\title{
AC Magnetic Field Sensing Using Continuous-Wave Optically Detected Magnetic Resonance of Nitrogen Vacancy Centers in Diamond
}
\author{Soya Saijo}
   \affiliation{
\keio
   }
\author{Yuichiro Matsuzaki}
   \affiliation{
\NTT
   }
\author{Shiro Saito}
   \affiliation{
 \NTT
   }
\author{Ikuya Hanano}
   \affiliation{
\keio
   }
\author{Hideyuki Watanabe}
   \affiliation{
\AIST
   }
\author{Norikazu Mizuochi}
   \affiliation{
\KYOTO
   }
\author{Junko Ishi-Hayase }
  \affiliation{
\keio
   }

\begin{abstract}
Nitrogen-vacancy (NV) centers in diamond are considered
sensors for detecting magnetic fields.
Pulsed optically detected magnetic resonance (ODMR) is typically used to detect AC magnetic fields;
however, this technique can only be implemented after careful calibration that involves
aligning an external static magnetic field, measuring continuous-wave (CW) ODMR,
determining the Rabi frequency, and setting the microwave phase.
In contrast, CW-ODMR can be simply implemented by continuous application of green
CW laser and a microwave filed.
In this letter, we report a method that uses NV centers and CW-ODMR to detect AC magnetic fields.
Unlike conventional methods that use NV centers to detect AC magnetic fields,
the proposed method requires neither a pulse sequence nor an externally applied DC magnetic field;
this greatly simplifies the procedure and apparatus needed to implement this method.
This method provides a sensitivity of
$2.5 \ \othermu \mathrm{T}/\sqrt{\mathrm{Hz}}$ at room temperature.
Thus, this simple alternative to existing AC magnetic field sensors paves
the way for a practical and feasible quantum sensor.
\end{abstract}

\maketitle

Nitrogen-vacancy (NV) centers in diamond can be used as sensitive room temperature
magnetic field sensors with submicron spatial resolution
\cite{Wolf2015a,Balasubramanian2008,Taylor2008}.
Many variations of NV center magnetic field sensors exist,
such as vector magnetic field sensing via confocal microscopy\cite{Maertz2010,Schoenfeld2011},
rapid imaging using charge coupled device (CCD) arrays\cite{Pham2011,LeSage2013a,Steinert2013,DeVience2015,Glenn2015},
and atomic force microscopy, which provides nanoscale imaging
using a nanodiamond or a diamond nanopillar tip\cite{Maletinsky2012,Rondin2012,Tetienne2014,Appel2016,Chang2016}.

To demonstrate a magnetic field sensor, either a pulsed ODMR technique or CW-ODMR technique may be used.
The pulsed ODMR technique has been applied to detect both AC and DC magnetic fields
\cite{Degen2008a,Maze2008,Taylor2008}.
Although the pulsed ODMR technique provides better sensitivity than the CW-OMDR technique,
it requires careful calibration before it can detect a magnetic field.
This calibration typically involves aligning an external static magnetic field,
measuring CW-ODMR, observing Rabi oscillations (to determine the Rabi frequency),
controlling the microwave phase, and constructing a pulse sequence.
Pulsed ODMR detects high frequency magnetic fields by narrowing the pulse interval,
which is technically possible,
but this method is not easy from the viewpoint of the coherence time of NV center and the cost of the high speed control device.

In contrast, the CW-ODMR technique can be used to detect DC magnetic fields or
low frequency (e.g., kHz) AC magnetic fields, and it is a more convenient technique
because it only requires the continuous application of microwaves and an optical laser.
Although the sensitivity of CW-ODMR is currently lesser than that of the pulsed ODMR,
its simple experimental requirements have led many researchers to use it for practical
and feasible magnetic field measurements.

To extend the applications of CW-ODMR method, using CW-ODMR with NV centers in diamond,
we developed a method to measure AC magnetic fields up to MHz frequencies.
Already, the CW-ODMR method is being used to measure AC magnetic fields in the kHz frequency range;
in this technique, magnetic fields are applied to exploit the two level nature of NV centers\cite{Jensen2014,Ahmadi2017,Tashima2017}.
In contrast, in the present work, we use the spin-1 properties of NV centers to measure AC magnetic fields with MHz frequencies.
Three energy eigenstates exist in the ground state manifold of NV centers,
all of which are used for magnetic field sensing.
The lowest energy eigenstate $|0\rangle $ is about 2.87 GHz below the two higher energy eigenstates,
which themselves have an energy difference in the order of MHz.
The idea behind the proposed method is to use this MHz transition frequency to detect AC magnetic fields
while the lowest energy eigenstate is continuously excited by the continuous microwave radiation.

Note that the CW-ODMR technique is compatible with CCD based techniques that have slow camera frame rate.
Since CCD cameras detect a wide field, the magnetic field information in diamond may be collected over a wide area in a single measurement.
This allows the magnetic field distribution to be rapidly acquired because, unlike other techniques,
the magnetic field in diamond does not need to be measured point by point.
However, a potential problem of the CCD based scheme is the slow camera operation time (from 100 Hz to 1 kHz).
Since the pulse repetition rate exceeds a few MHz for typical AC magnetic field sensing,
sophisticated techniques such as the use of optical shutters are required\cite{Pham2011,DeVience2015}.
Conversely, because the CW-ODMR technique does not invoke such fast operations,
in our AC magnetic field sensors provides a way to adopt the CCD based technique with much more simple experimental setup.
Since the CCD based setup can increase the measurement volume of NV centers,
the signal from the NV is enhanced and highly sensitive sensing becomes possible.
Furthermore, no external static magnetic field is used, the measurement volume can be increased without concern for the uniformity of the static magnetic field.

\begin{figure}[b]
\begin{center}
\includegraphics[width=0.6\linewidth]{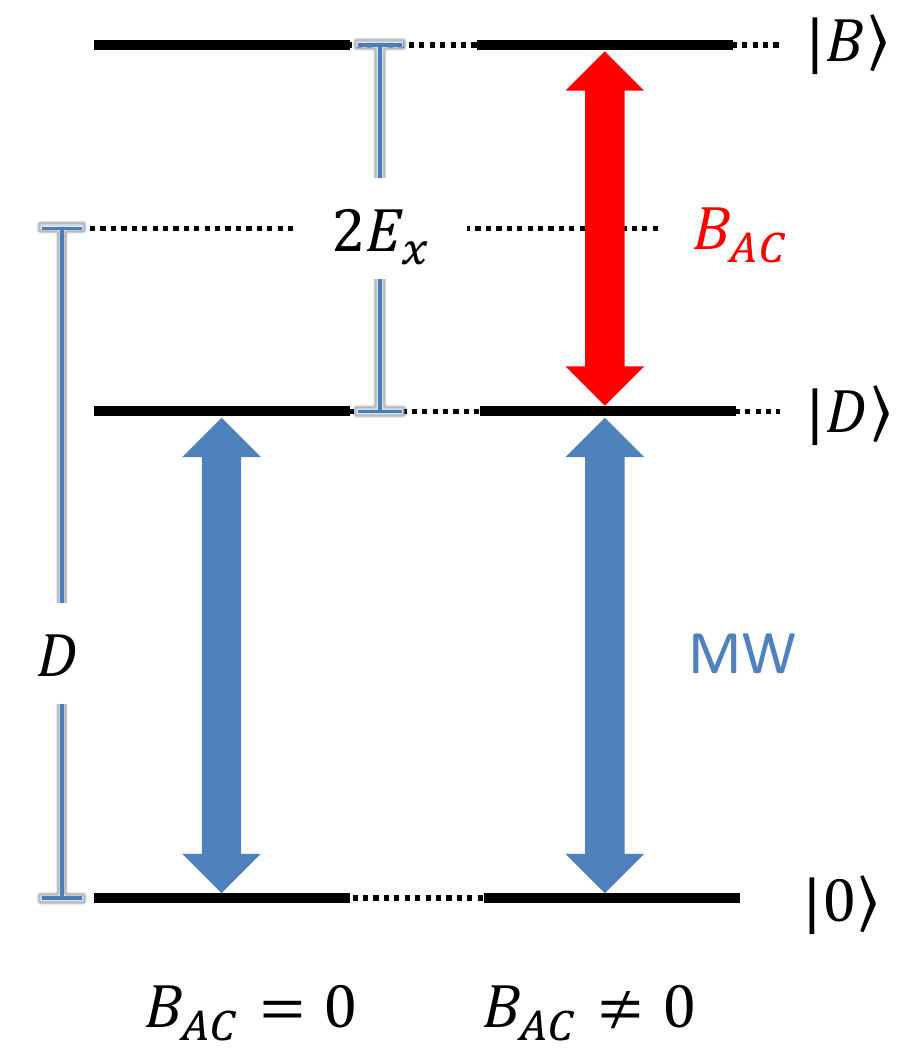}
\caption{
Energy level diagram of the NV center in diamond.
To drive these states, we apply both a GHz frequency microwave field and a MHz frequency AC magnetic field.
}
\label{energy}
\end{center}
\end{figure}
We start by explaining the theory behind both the conventional methods and the proposed method.
The Hamiltonian of an NV center with no external magnetic field is given as follows.
\begin{equation}
H_{\mathrm{NV} }=  D{ \hat { S }  }_{ z }^{ 2 }
+ { E }_{ x } \left( { \hat { S }  }_{ x }^{ 2 }-{ \hat { S }  }_{ y }^{ 2 } \right)
+ { E }_{ y } \left( { \hat { S }  }_{ x }{ \hat { S }  }_{ y } - { \hat { S }  }_{ y }{ \hat { S }  }_{ x } \right)
\label{HNV}
\end{equation}
where $ \hat { S } $ is a spin-1 operator for electron spin,
$ D $ is the zero field splitting,
and ${ E }_{ x }({ E }_{ y })$ is the strain in the x(y) direction.
Without loss of generality, we set $E_y=0$ by defining
the $x$ axis to pass through the NV center in the direction of the strain.
Throughout this letter, we set $\hbar=1$.
The ground state is $|0\rangle $,
and we define the two higher energy eigenstates as
$\ket{ D } =    \left( \ket{ 1 } - \ket{ -1 } \right)/\sqrt { 2 }$ and
$\ket{ B }  =  \left( \ket{ 1 } + \ket{ -1 } \right)/\sqrt { 2 } $
with eigenenergies $D-E_x$ and $D+E_x$, respectively.
With zero external magnetic field, two dips appear around 2.87 GHz in CW-ODMR,
which is indicative of externally driven transitions from the ground state
$|0\rangle $ to higher energy eigenstates such as $|B\rangle $
or $|D\rangle $\cite{Fang2013,Zhu2014,Matsuzaki2016}.

We now consider the dynamics of NV centers
when both the microwave field and the target AC magnetic field are present.
With these external fields, the Hamiltonian of the NV center takes the form
 $ H = H_{\mathrm{NV}}+H_{\mathrm{ex}}$, where $H_{\mathrm{ex}}$ is given by
\begin{equation}
H_{\mathrm{ex}} = \sum_{j=x,y,z}{{\gamma}_eB}^{(j)}_{\mathrm{mw}} { \hat { S }
 }_{ j }\cos { \left ({ \omega  }_{ \mathrm {mw} }t \right)  } +
 {{\gamma}_eB}^{(j)}_{\mathrm AC } { \hat { S }  }_{ j } \cos {
 \left({\omega}_{\mathrm {AC}}t \right) }
\label{Hex}
\end{equation}
where ${\gamma}_e$ is the gyromagnetic ratio of the electron spin,
${ B }_{ \mathrm{mw} }({ B }_{ \mathrm{AC} })$ is the microwave field (target AC magnetic field) amplitude,
and ${ \omega  }_{ \mathrm{mw} }({ \omega  }_{ \mathrm {AC} })$ is the frequency of the driving microwave field (target AC magnetic field).
We assume that ${ \omega  }_{ \mathrm {mw} }$ is in the order of GHz, whereas ${ \omega  }_{ \mathrm {AC} }$ is in the order of MHz.
In a rotating frame defined by $ U $, the effective Hamiltonian becomes $ H' = U H \cre{ U } - iU\ddt \cre{ U }$.
By considering $ U={ e }^{ i { \omega  }_{ \mathrm {mw} } t { \hat { S }  }_{ z }^{ 2 } }$ and using the rotating wave approximation,
we obtain the following Hamiltonian.
\begin{align}
{ H } \simeq& \left( D - { \omega  }_{ \mathrm{mw} } \right) { \hat { S }  }_{ z }^{ 2 }
+  { E }_{ x } \left( { \hat { S }  }_{ x }^{ 2 }-{ \hat { S }  }_{ y }^{ 2 } \right)  \nonumber \\
&+ \frac { {{\gamma}_eB}^{(x)}_{\mathrm{mw}} }{ 2 } { \hat { S }  }_{ x }
+ \frac { {{\gamma}_eB}^{(y)}_{\mathrm{mw}} }{ 2 }  { \hat { S }  }_{ y }
+ {{\gamma}_eB}^{(z)}_{\mathrm{AC}} { \hat { S }  }_{ z } \cos {\left ({ \omega  }_{ \mathrm{AC} }t \right) }  \\
=&\left( D + E _ x- { \omega  }_{ \mathrm{mw} } \right) \ket{ B } \bra{ B }
+ \left( D - E _ x- { \omega  }_{ \mathrm{mw} } \right) \ket{ D } \bra{ D } \nonumber \\
&+ \frac { { {\gamma}_eB }^{ ( x ) }_{\mathrm{mw} } }{ 2 } \left( \ket{ B }\bra{ 0 } + \ket{ 0 } \bra{ B }\right)
-\frac { i{ {\gamma}_eB }^{ (y) }_{\mathrm{mw}} }{ 2 } \left( \ket{ D } \bra{ 0 } - { \ket{ 0 } } \bra{ D } \right)\nonumber \\
&+ {{\gamma}_eB}^{(z)}_{\mathrm{AC}}  \left( \ket{ B } \bra{ D } + \ket{ D } \bra{ B } \right) \cos { \left ({ \omega  }_{ \mathrm{AC} }t \right)  }
\label{H1}
\end{align}
In a different rotating frame defined by
$U'=e^{i\frac{\omega _{\mathrm{AC}} }{2}
\left( { \hat { S } }_{ x }^{ 2 }-{ \hat { S } }_{ y }^{ 2 } \right)}$,
we obtain the following Hamiltonian.
\begin{align}
H=&\left( D+E_{x} - { \omega  }_{ \mathrm{mw} } -\frac{1}{2}{\omega}_{\mathrm{AC}}\right) \ket{ B } \bra{ B } \nonumber \\
&+\left( { D } - E _ x - { \omega  }_{ \mathrm{mw} } +\frac{1}{2}{\omega}_{\mathrm{AC}}\right) \ket{ D } \bra{ D } \nonumber \\
&+\frac{{{\gamma}_eB}^{(x)}_{\mathrm{mw}} }{ 2 } \left( e^{\frac{i\omega_{\mathrm{AC}}t}{2}} \ket{ B } \bra{ 0 }
+ e^{\frac{-i\omega_{\mathrm{AC}}t}{2}} \ket{ 0 } \bra{ B } \right)\nonumber \\
&- \frac { i{{\gamma}_eB}^{(y)}_{\mathrm{mw}} }{ 2 } \left( e^{\frac{-i\omega_{\mathrm{AC}}t}{2}} \ket{ D } \bra{ 0 }
- e^{\frac{i\omega _{\mathrm{AC}}t}{2}} \ket{ 0 } \bra{ D } \right)\nonumber \\
&+ \frac{1}{2}{{\gamma}_eB}^{(z)}_{\mathrm{AC}}  \left( \ket{ B } \bra{ D } + \ket{ D }  \bra{ B } \right)
\label{H2}
\end{align}
where we have again used the rotating wave approximation.

Importantly, an AC magnetic field in the $z$ direction
[the fifth term in the Hamiltonian (\ref{H2})] induces the transition between $ \ket{ B }$ and $ \ket{ D } $
when the frequency of the field in resonance with the energy difference between $ \ket{ B }$ and $ \ket{ D } $.
Without the AC magnetic field, we can only induce transitions between the ground state $ \ket{ 0 } $
and the bright (dark) state $ \ket{ B } $ [$ \ket{ D } $] via the microwave radiation with a
frequency of ${ \omega  }_{ \mathrm{mw} } \simeq D+E_x$ (${ \omega  }_{ \mathrm{mw} } \simeq D-E_x$)
in the conventional CW-ODMR setup.
$ \ket{ B }$ and $ \ket{ D } $ the applied AC magnetic field and the microwave field
can drive transitions from the ground state to the bright and dark states.
Thus, the results of CW-ODMR with an applied AC magnetic field should differ from t
hose of CW-ODMR without any AC magnetic field.

We now quantify the change in the CW-ODMR signal that occurs because of the AC
magnetic fields first focusing on the transition induced between
$|B\rangle$ and $|D\rangle$ by the AC magnetic field with a
frequency of $\omega _{\mathrm{AC}}=2E_x$.
We assume a weak amplitude for the AC magnetic field so that
we can use time dependent perturbation theory.
In the interaction picture, we obtain
\begin{align}
H _I= \frac { {{\gamma}_eB}^{(x)}_{\mathrm{mw}} }{ 2\sqrt{2} }
&e^{i(D-\omega_{\mathrm{mw}}+\frac{{\gamma}_eB_{\mathrm{AC}}^{(z)}}{2}+E_x)t}{ \ket{ 1 } } \bra{ 0 } \nonumber \\
&+e^{i(D-\omega_{\mathrm{mw}}-\frac{{\gamma}_eB^{(z)}_{\mathrm{AC}}}{2}+E_x)t} \ket{ -1 }  \bra{ 0 } + {\mathrm{hc}})\nonumber \\
-\frac { i{{\gamma}_eB}^{(y)}_{\mathrm{mw}} }{ 2 \sqrt{2}}
&(e^{i(D-\omega_{\mathrm{mw}}+\frac{{\gamma}_eB_{\mathrm{AC}}^{(z)}}{2}-E_x)t} \ket{ 1 }  \bra{ 0 }\nonumber \\
&-e^{i(D-\omega_{\mathrm{mw}}-\frac{{\gamma}_eB^{(z)}_{\mathrm{AC} }}{2}-E_{x})t} \ket{ -1 } \bra{ 0 }+ {\mathrm hc})
\label{HI}
\end{align}
By using Fermi's golden rule, we can show that the transition from the ground state to the higher energy eigenstates
$\omega_{\mathrm{mw}}\simeq D\pm {\gamma}_eB^{(z)}_{\mathrm{AC}}/ 2 \pm E_x$ (Fig. \ref{energy}).
Thus, applying an external AC magnetic field changes the dip structure in CW-ODMR.

We now describe the details of the diamond sample used in our experiment.
We used an ensemble of NV centers in a diamond film on a (001) electronic grade substrate.
The isotopically purified $^{ 12 }{\mathrm{C} }$ diamond film ($\left[ ^{ 12 }{\mathrm{C} } \right] =99.999\ \%$)
was grown using nitrogen doped microwave plasma assisted chemical vapor deposition.
To both increase the NV center density and improve the coherence time\cite{Kleinsasser2016},
the sample was irradiated with ion doses of ${ 10 }^{ 12 }\  { \mathrm{cm} }^{ -3 }$ with 15 keV He$^{+}$ ions and
was annealed for 24 h in vacuum at 800 ${\ }^\circ \mathrm{C}$.
The NV density was estimated to be of the order of $ { 10 }^{ 15} \ {\mathrm{cm} }^{ -3 }$.

Now, we explain the experiment of sensing an AC magnetic field using CW-ODMR.
For these experiments, we used a homebuilt system for confocal laser scanning microscopy with a spatial resolution of 400 nm.
The diamond sample was positioned above the antenna\cite{Sasaki2017} used to emit the microwave radiation.
A 30 $\othermu$m diameter copper wire is placed in contact with the sample surface to apply the target AC magnetic field,
which is detected by measuring the difference in the CW-ODMR spectrum.

\begin{figure}[t]
\centering
\includegraphics[width=\linewidth]{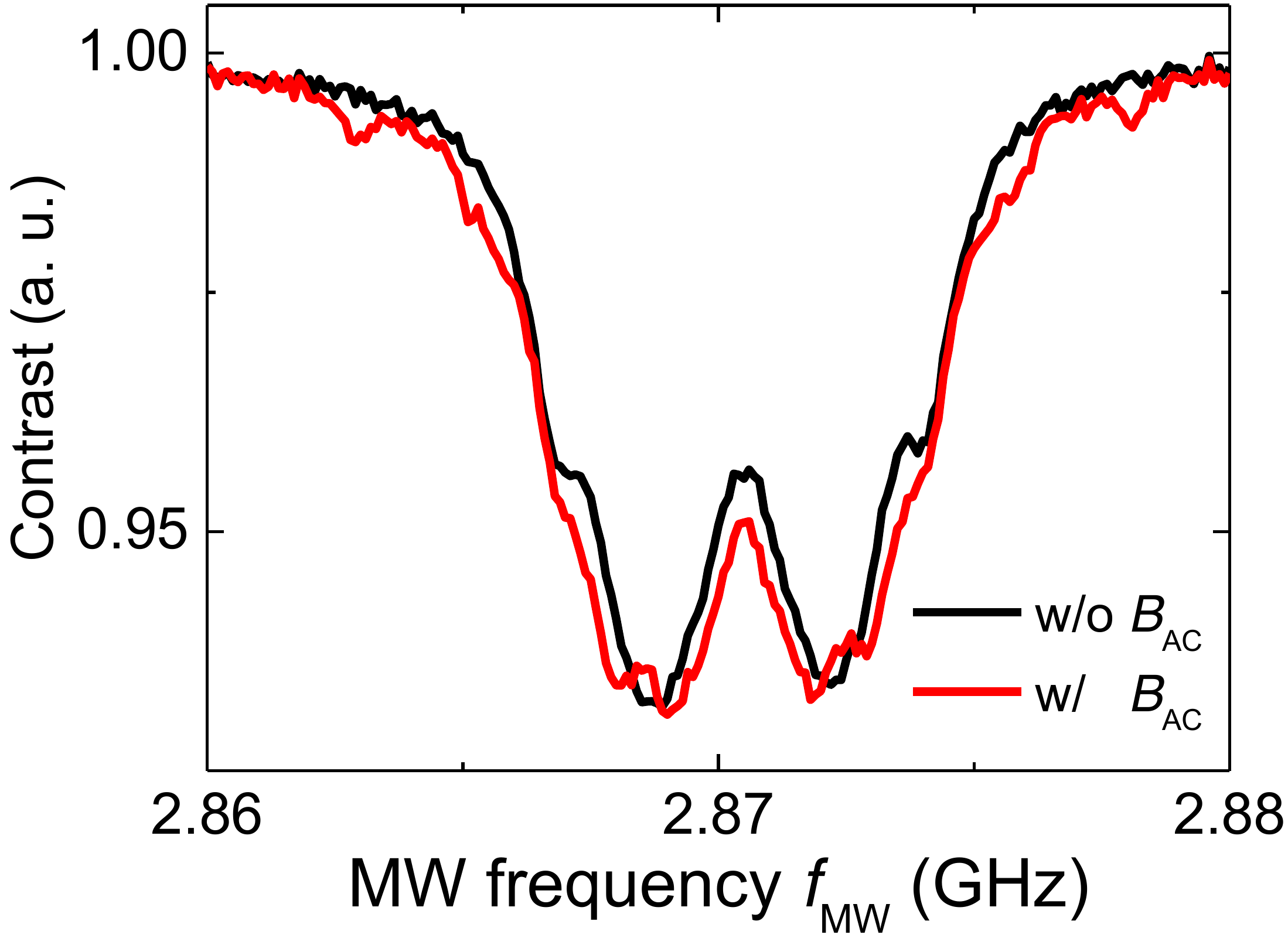}
\caption{ODMR with and without an AC magnetic field
 (${B}_{\mathrm {AC}} \cos 2\pi { f }_{ \mathrm{AC} }$) for
  ${B}_{\mathrm {AC}} =
 7.7 \ \othermu \mathrm{T}$ and
  ${ f }_{ \mathrm {AC} } = 4 \ \mathrm{MHz}$.
  Here, no DC magnetic field is applied.
 The application of an AC magnetic field changes the ODMR spectrum.}
\label{ODMR}
\end{figure}
Figure \ref{ODMR} shows the signal from the conventional CW-ODMR technique (with no external AC magnetic field);
the resonance frequency is split by about 4 MHz because of a local magnetic field and strain from impurities in diamond\cite{Zhu2014,Matsuzaki2016}.
This splitting gives the energy difference between $ \ket{ B } $ and $ \ket{ D } $.
ODMR is performed by applying an AC magnetic field with a frequency of
${ f }_{ \mathrm{AC} } =  { \omega }_{ \mathrm{AC} } / 2 \pi = 4 \ \mathrm{MHz}$
and a magnetic field amplitude ${ B }_{ \mathrm{AC} } = 7.7 \ \othermu \mathrm{T}$ to induce
transitions between $ \ket{ B } $ and $ \ket{ D } $.
The result in Fig. \ref{ODMR} shows the
difference in the spectrum due to the external AC magnetic field,
and it demonstrates the detection of the external AC magnetic fields by CW-ODMR.

Since we use the resonance between $ \ket{ B } $ and $ \ket{ D } $,
the frequency band in which AC magnetic fields may be detected is determined by the width of the split in the CW-ODMR spectrum.

As noted in Eq. (\ref{H2}), this splitting may be increased by applying an electric field\cite{Iwasaki2017},
which provides a way to determine which frequency of the AC magnetic field will be detected.
Straightforward calculations show that the detectable frequencies range from hundreds of kHz to hundreds of MHz.
The lower limit is determined by the resonance linewidth of the ODMR spectrum with 200 kHz currently being the minimum linewidth\cite{Kleinsasser2016}.
However, the upper limit is determined by the breakdown field in diamond.
The splitting width due to the Stark effect is given as $2R{E}_{x}$, so the maximum splitting width is 340 MHz
because the Stark shift constant $R=$ 17 Hz cm/V\cite{VanOort1990} and the breakdown electric field of diamond $E=$ 10 MV/m\cite{Collins1994}.

\begin{figure}[b]
\centering
\includegraphics[width=\linewidth]{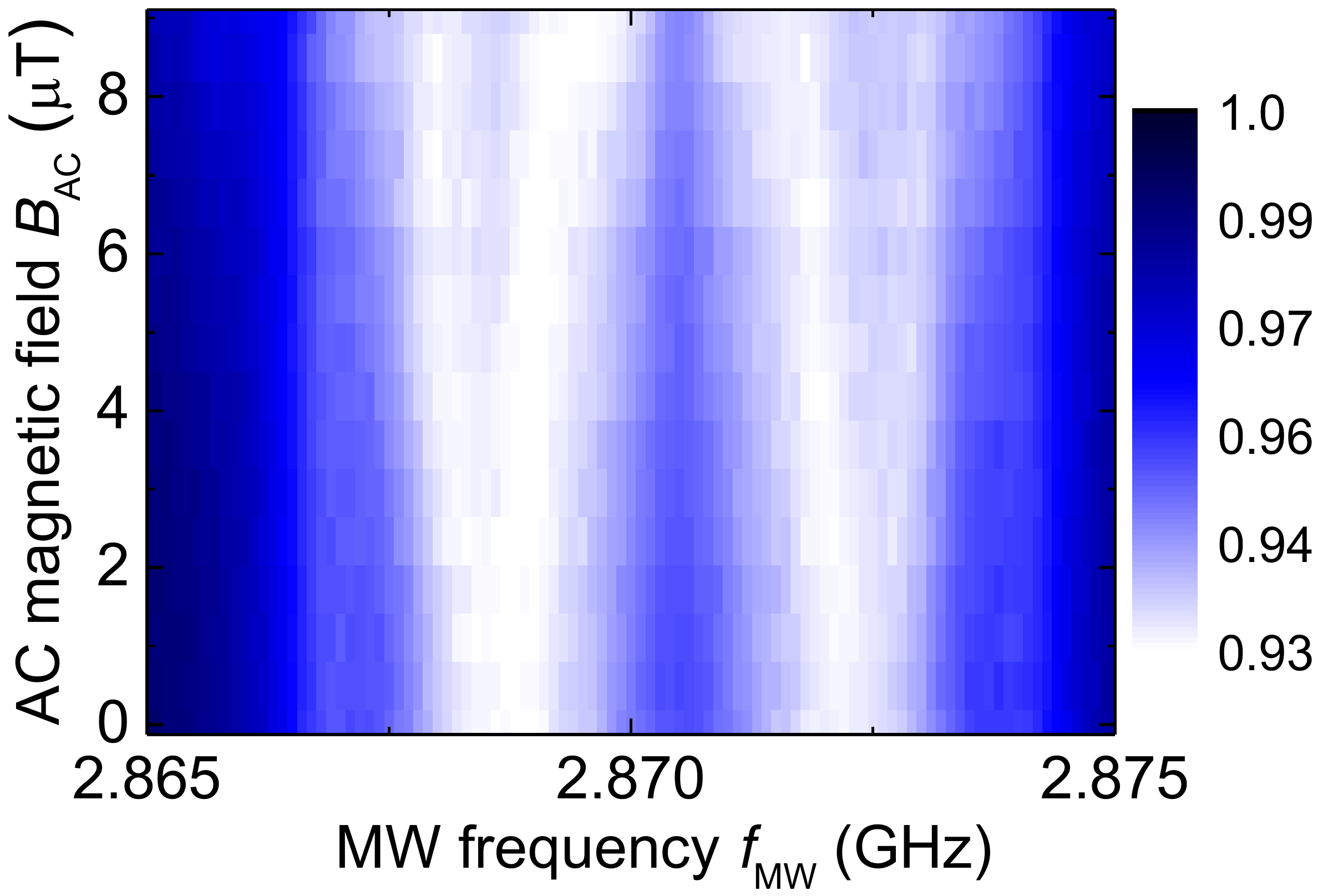}
\caption{
ODMR spectrum with applied AC magnetic fields.
Here, the x and y axes denote the frequency of the microwave and the amplitude of the AC magnetic fields,
respectively. Four resonances are observed with the applied AC magnetic fields.
}
\label{ODMR_B}
\end{figure}

Next, we measure ODMR dependence on the amplitude of the AC magnetic field.
The ODMR spectrum with various AC magnetic field amplitudes  is shown in
Fig. \ref{ODMR_B} where we set ${ f }_{ \mathrm{AC} } = 4 \ \mathrm{MHz}$.
While two resonances are
 observed without applying the AC magnetic fields, the resonance is
 split into four lines, and the split becomes larger as we increase the
 amplitude of the AC magnetic fields. This is consistent with our derived
 formula that represents the resonance $\omega_{\mathrm{mw}}\simeq D\pm
{\gamma}_eB^{(z)}_{\mathrm{AC}}/2\pm E_x$.

\begin{figure}
\centering
\includegraphics[width=\linewidth]{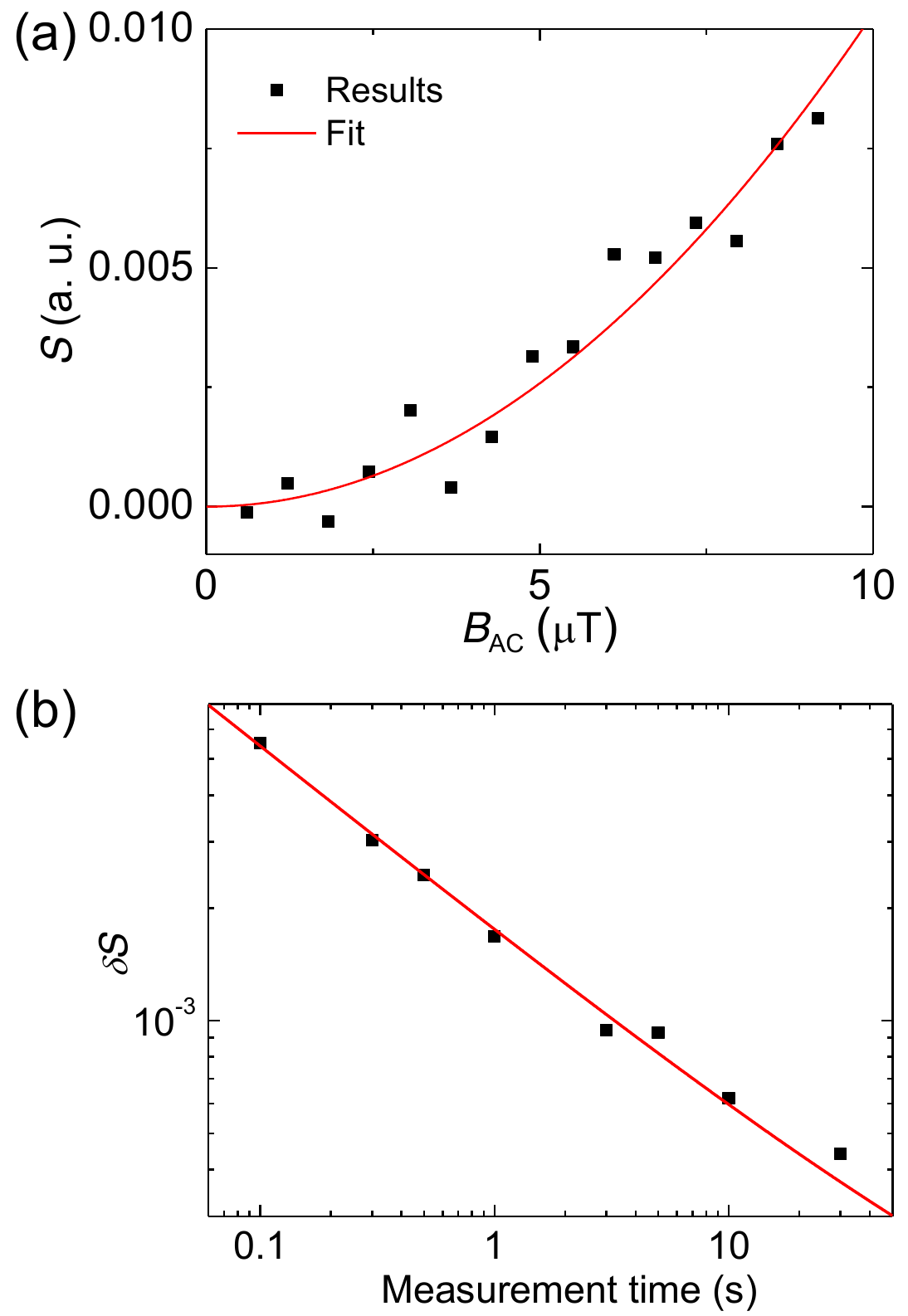}
\caption{
 Measured signal and standard deviation of the ODMR
 signal. (a) ODMR signal plotted against the amplitude of the AC magnetic field.
 The data are fit to the function $S-S_0=a
 (B_{\mathrm{AC}})^2$, where $a$ is a fitting parameter and $S_0$ is
 the offset. The fit shows that the signal is quadratic in the amplitude of the AC magnetic field.
 (b) Standard deviation $\delta S$ of the ODMR signal as a function of measurement time.
}
\label{strength}
\end{figure}

In Fig. \ref{strength}(a), we plot the ODMR signal against the amplitude of
 the AC magnetic field, with the microwave frequency ${ f }_{ \mathrm{mw} } =  { \omega
 }_{ \mathrm{mw} } / 2 \pi =2.86887\ \mathrm GHz$, which is one of the two
 resonance frequencies obtained with no external AC magnetic field.
 The signal quadratically depends on the amplitude, which can be quantitatively understood as follows.
 We represent the two resonances around $D-E_x$ as the sum of two Lorentzian functions:
 $F(B)=\sum_{j=\pm 1}^{}1/(({ f }_{ \mathrm{mw} } -(D+j{\gamma}_eB^{(z)}_{\mathrm{AC}} /2 - E_x) )^2+\gamma ^2)$,
 where $\gamma$ is the linewidth. In this case, we obtain
 $dF(B)/dB\propto B$ for small $B$, which this shows the
 quadratic dependence.

To estimate the magnetic sensitivity
$\delta { B }_{\mathrm{AC}}$ from the experimental results, we must determine
the signal fluctuation $\delta S$ where $S$ corresponds to the
photoluminescence in ODMR \cite{Taylor2008}.
As shown in Fig. \ref{strength}(b), the fluctuation decreases with the relation $\sqrt{ T }$, where $T$ is the measurement time.
From these experimental observations, we estimate the
sensitivity of the method for detecting AC magnetic fields to be $2.5 \ \othermu \mathrm{T}/\sqrt{\mathrm{Hz}}$.
Note that the sensitivity could be improved by using a NV center in diamond with a narrow linewidth\cite{Kleinsasser2016}
and with an almost perfect preferential orientation of the axis\cite{Michl2014a,Fukui2014,Ozawa2017} and
a high density of the NV centers\cite{Acosta2009,Kleinsasser2016,Ozawa2017}.
 In fact, by using the parameters reported in \cite{Ozawa2017},
 we estimate that the proposed method would have a sensitivity of $50\ \mathrm {nT}/\sqrt{\mathrm {Hz}}$.

In conclusion, we report a method to detect MHz frequency AC magnetic fields that uses CW-ODMR
by exploiting NV centers in diamond. By simply applying a continuous microwave field and optical laser irradiation,
the method provides a sensitivity of $2.5 \ \othermu \mathrm {T}/\sqrt{\mathrm {Hz}}$
at room temperature.
The experimental setup is very simple because it requires neither an external DC magnetic
field nor a pulse sequence. These results pave the way to realize a practical and feasible AC magnetic field sensor.

We thank H. Toida and
K. Kakuyanagi for helpful discussions.
This work
 was supported by JSPS KAKENHI Grant No. 15K17732. This
 work was also supported by MEXT KAKENHI Grants No.
 15H05868, No. 15H05870, No. 15H03996, No. 26220602,
 and No. 26249108. This work was also supported by the
 Advanced Photon Science Alliance (APSA), JSPS Core-to-
 Core Program FY2013 Projects No.2, and Spin-NRJ.

\end{document}